\documentclass[prd, preprint, 12pt]{revtex4-1}

\usepackage{amsmath}
\usepackage{amssymb}
\usepackage{setspace}
\usepackage{graphicx}
\usepackage{natbib}
\usepackage{float}

\begin{document}
 
 %

\begin{center}
 { \large {\bf A Duality Between Curvature and Torsion }}


\vskip 0.4 in

{\large{\bf Swanand Khanapurkar$^{*\dagger}$ and Tejinder P.  Singh$^\dagger$}}

\medskip
{\it $^{*}$Indian Institute of Science Education and Research, Pune 411008, India }\\  
{\it $^\dagger$Tata Institute of Fundamental Research,}
{\it Homi Bhabha Road, Mumbai 400005, India}\\
\bigskip
{\tt email:  swanand.khanapurkar@students.iiserpune.ac.in, tpsingh@tifr.res.in}\\

\end{center}

\bigskip
\bigskip

\centerline{\bf ABSTRACT}
\noindent Compton wavelength and Schwarzschild radius are considered here  as limiting cases of a unified length scale. Using this length, it is shown that the Dirac equation and the Einstein equations for a point mass are limiting cases of an underlying theory which includes torsion. We show that in this underlying theory the gravitational interaction between small masses is weaker than in Newtonian gravity. We explain as to why the Kerr-Newman black hole and the electron both have the same non-classical gyromagnetic ratio.  We propose a duality between curvature and torsion and show that general relativity and teleparallel gravity are respectively the large mass and small mass limit of the ECSK theory. We demonstrate that small scale effects of torsion can be tested with current technology.

\bigskip
\noindent 
\noindent 

\vskip 1 in

\centerline{March 29, 2018}

\bigskip

\centerline{Essay written for the Gravity Research Foundation 2018 Awards for Essays on Gravitation}
\bigskip

\centerline {{\bf Corresponding Author:} Tejinder P. Singh}
\bigskip

\newpage

\setstretch{1.3}

\bigskip

\noindent It is often stated that when the special theory of relativity is generalized to include general coordinate transformations and non-inertial observers,  the equivalence principle leads to the general theory of relativity. There is a little oddity in this statement. The symmetry group of special relativity is the Poincar\'e group, which includes the Lorentz group as well as 
space-time translations. However, the local symmetry group of general relativity is not the Poincar\'e group, but the Lorentz group. Translations are not included. If one makes the rather natural assumption  that the local symmetry group of space-time should be the Poincar\'e group, one gets a Poincar\'e gauge theory of gravity, which includes torsion, besides curvature. The Lorentz group corresponds to curvature and mass-energy. Translations correspond to torsion and spin. An elegant example of Poincar\'e gauge gravity is the Einstein-Cartan-Sciama-Kibble [ECSK] theory \cite{Hehl}, which is a minimal generalisation of GR which incorporates torsion via the anti-symmetric part of the connection. Poincar\'e gauge gravity is closer in spirit to particle physics than GR is, because elementary particles are irreducible representations of the Poincar\'e group [not the Lorentz group], and are labelled by both mass and spin [not just mass], and spin couples naturally to torsion.

Why  do we then not observe torsion in the universe around us? In the ECSK theory, torsion vanishes outside of matter, so that in free space ECSK is the same as GR. Inside matter, say in a fluid of particles of mass $m$, torsion becomes comparable to curvature only at length scales smaller than the Einstein-Cartan radius $r_c = (\lambda_c L_{Pl}^2)^{1/3}$, and at densities higher than $m/r_c^3$, where $\lambda_c=\hbar/mc$ and $L_{Pl}$ are Compton wavelength and Planck length respectively. For nucleons, the Einstein-Cartan radius is about $10^{-27}$ cms, and the density above which torsion becomes important is about $10^{54}$ gms/cc \cite{Hehl}. These scales are beyond current technology, and since GR is in excellent agreement with observations, it is said that torsion can be safely neglected in today's universe. This is the conventional wisdom.

In this essay we question this conventional wisdom, by noting that there is a part of the mass parameter space where GR actually does not agree with experiments! For instance, GR claims to describe the dynamics and  space-time geometry of a particle of mass $m$ by the Schwarzschild space-time, for arbitrary values of $m$, including $m$ as small as the mass of the electron. But we know that the correct description of the electron is given by the Dirac equation, not by GR. One might of course object and say that GR is not intended for small masses. But how small is small? There is no mass scale in GR. Similarly, there is no mass scale in the Dirac equation, which claims to hold for all masses; and we know this claim contradicts the dynamics of large masses, which are correctly described by GR. To resolve this conflict between  GR and the Dirac equation, we introduce a mass scale, namely the Planck mass $m_{Pl}$, and propose a new theory, to which the Dirac equation and GR are small mass and large mass approximations respectively. Doing so compels us to incorporate torsion, and introduces an elegant symmetry and duality between torsion and curvature. Furthermore, we show that torsion is indeed important in the present universe, on microscopic scales of the order of Compton wavelength of elementary particles. 

We define a new length scale $L_{CS}$, dependent on $m$, and dubbed the Compton-Schwarzschild wavelength \cite{Carr}, having the property that $L_{CS}$ remains unchanged under the map $m\rightarrow m_{Pl}^2/4m$. For $m\ll m_{Pl}$, \; $L_{CS}=\lambda_c/2$; for $m\gg m_{Pl}$, \; $L_{CS}=2Gm/c^2$, and $L_{CS}$ has a minimum value $L_{Pl}$ at $m= m_{Pl}/2$ \cite{Singh1}. Using this length scale, we construct the following action for the interaction of a Dirac field $\psi$ with its own gravitational field \cite{Singh2}
\begin{equation}
\frac{L_{Pl}^2 }{\hbar} S =  \int d^4x \;\sqrt{-g} \left[ \frac{1}{8\pi} R\;  - \frac{1}{2}\; L_{CS}\; {\overline\psi}{\psi} \; + \; L_{CS}^2 \; \left\{\frac{i}{2}{\overline\psi}\gamma^{\mu}\; \nabla_\mu\psi  - \frac{i}{2} (\nabla_{\mu}\overline\psi)\gamma^{\mu}\psi\right\}
\; \right] 
\label{actualaction3}
\end{equation}
where the symbols have their usual meaning. For masses much smaller than Planck mass, assuming the gravitational field can be neglected, this action yields the Dirac equation on Minkowski space-time. For masses much larger than Planck mass, this action yields Einstein field equations for a point mass, assuming that the Dirac kinetic term can be neglected, and $\overline{\psi}\psi$ in the mass term can be replaced by a spatial delta-function [we justify this below]. Thus this action provides GR and the Dirac equation as limiting cases for a point mass. For an arbitrary value of the mass, the action yields the Einstein-Dirac equations, but now with the new coupling constant $L_{CS}$. This new coupling constant leads to an unexpected consequence. In the Einstein equations, for small masses, the  matter-gravity coupling constant $L_{CS}$ goes to $\hbar/2mc$, and no longer depends on Newton's constant $G$. The gravitational field on the left hand side can be expected to vanish [no $G$]. However the matter energy-momentum on the right hand side is non-vanishing. To overcome this contradiction, a natural resolution is to include torsion: for small masses the left hand side is dominated by the contribution of torsion to curvature, and the right hand side is dominated by the [now to be added] contribution of spin angular momentum density to the total matter energy-momentum-spin tensor. Torsion now couples to spin through $\hbar$ (not $G$), which seems rather natural.
Hence we generalize the above action to minimally include torsion via the antisymmetric connection in the covariant derivative, just as in the ECSK theory. Variation of the action with respect to the metric, the torsion (more precisely, the contortion), and the Dirac state yields the Einstein-Cartan-Dirac [ECD] equations
but now with $L_{CS}$ instead of $L_{Pl}$ and $\lambda_c$:
\begin{equation}
G^{ij} = 8\pi \; L_{CS}^2 \;  \frac{1}{\hbar c}\; \Sigma^{ij}
\label{TPS1}
\end{equation}
\begin{equation}
T^{ijk} =8\pi \;  L_{CS}^2 \;  \frac{1}{\hbar c} \; \tau^{ijk}
 \label{TPS2}
 \end{equation}
 \begin{equation}
 i\gamma^{\alpha} \nabla^{\{\}}_{\alpha}\psi + \frac{3}{8} L_{CS}^2 \; (\overline{\psi}\gamma_5 \gamma^{\alpha}\psi) \gamma_5 \gamma_{\alpha}\psi - \frac{1}{2 L_{CS}} \psi =0
 \label{TPS3}
 \end{equation}
 In Eqn. (\ref{TPS1}) the $G^{ij}$ on the left hand side is the asymmetric Einstein tensor built from the asymmetric connection, which in turn is made from the symmetric Levi-Civita connection and the antisymmetric connection, i.e. the torsion tensor. The $\Sigma^{ij}$ on the right hand side is the asymmetric canonical total energy momentum tensor, which is made from the symmetric matter stress-energy-momentum, and from the spin angular momentum tensor. In Eqn. (\ref{TPS2}), the so-called 
 modified torsion $T^{ijk}$ is the traceless part of the torsion tensor, and it is algebraically related to the $\tau^{ijk}$ on the right, which is the spin angular momentum tensor, which results from varying the matter Lagrangian with respect to the contortion tensor. Thus, torsion couples to spin in a very natural manner. Eqn (\ref{TPS3}) is the non-linear Dirac equation, also known as the Hehl-Datta equation \cite{HD}, with the non-linear term coming because of the torsion dependent part of the covariant derivative, and because torsion can be algebraically related to spin, which in turn is expressed in terms of the Dirac state. This system of equations is the same as the standard ECD equations, except that in Eqns. (\ref{TPS1}) and (\ref{TPS2}) $L_{Pl}$ is replaced by $L_{CS}$, and in Eqn. (\ref{TPS3}) the $L_{Pl}$ in the nonlinear term, and  the $\lambda_c$ in the mass term are both replaced by $L_{CS}$. These equations hold for all values of the mass $m$, and it  is then only natural that the coupling constant should be $L_{CS}$, instead of $L_{Pl}$ and $\lambda_c$, for why should the latter two appear in the ECD equations for a large mass? 
 
 For $m\gg m_{Pl}$, torsion and spin are negligible, and these equations reduce to Einstein equations for a point mass. For $m\ll m_{Pl}$, spin dominates mass, and a novelty emerges, which we describe later below. From the structure of these equations it is evident that the Einstein-Cartan radius, where torsion becomes important, and which was earlier $r_c = (\lambda_c L_{Pl}^2)^{1/3}$, is now simply $L_{CS}$. This is of the order of Compton wavelength, for elementary particles, and it would be worthwhile to investigate what effect torsion has on these and smaller scales. It is worth noting that known experimental and theoretical bounds on torsion which have been investigated thus far all seem to be on scales comparable to or larger than Compton wavelength \cite{Shapiro}. One now needs to look into possible effects on smaller length scales. 
 
 Because of $L_{CS}$, these equations have some important consequences and predictions, which we now describe. Firstly, there is the curious fact that the Kerr-Newman black hole, despite being a classical object, has a non-classical gyromagnetic ratio $\gamma$, equal to that of the Dirac electron, both being twice the classical value $q/2m$, where $q$ is the electric charge of the object \cite{Newman}. This profound fact, for which there is no explanation, strongly hints at some possible connection between Dirac fermions on the one hand, and Kerr-Newman black holes on the other. In our work, since a Dirac fermion of mass $m_q$ and its dual black hole of mass $m_h = m_{Pl}^2 / 4m_q$ are both described by the same $L_{CS}$, and since the same set of field equations (\ref{TPS1}-\ref{TPS3}) describe both Dirac fermions and black holes, an explanation becomes possible. For a Kerr-Newman black hole of mass $m_h$ and charge $q_h$  we can write
 \begin{equation}
  \gamma = 2. q_h/2m_h = \frac{2q_h}{m_{Pl}} \frac{L_{Pl}}{L_{CS}} = \frac{2q_h}{m_{Pl}}\frac{L_{Pl}}{\hbar/2m_q c}
  =  2. \frac{q_h.4m_q^2}{2m_q}.\frac{cL_{Pl}}{\hbar m_{Pl}} = 2. \frac{q_h.4Gm_q^2/\hbar c}{2m_q} = 2. q/2m_q
  \end{equation}
  where $q=q_h.\hbar c/4Gm_h^2$.
  Thus the gyromagnetic ratio of a black hole of mass $m_h$ and charge $q_h$ is the same as that of a Dirac fermion with dual mass $m_q$ and dual charge $q= q_h.\hbar c/4Gm_h^2$. This explains why the Kerr-Newman black hole naturally has a g-ratio twice the classical value.
  
  We next consider the non-relativistic limit of the ECD equations (\ref{TPS1}-\ref{TPS3}) by substituting therein the following series expansion for the metric and the quantum state, in terms of the expansion parameter $\sqrt{\hbar}/c$ \cite{Swanand1, Swanand2}:
 \begin{equation}\label{eq:Spinor_ansatz_NRlimit}
	\psi(\textbf{r},t) = e^{\frac{ic^2}{\hbar}S(\textbf{r},t)}\sum_{n=0}^{\infty}\bigg(\frac{\sqrt{\hbar}}{c}\bigg)^n a_n(\textbf{r},t)
	; \qquad 
	g_{\mu\nu}({\bf r}, t) = \eta_{\mu\nu} + \sum_{n = 1}^{\infty} \bigg{(}\frac{\sqrt{\hbar}}{c} \bigg{)}^n g_{\mu\nu}^{[n]}({\bf r}, t)
	\end{equation}  
	Since the dependence of $L_{CS}$ on $c$ is different in the large mass limit and in the small mass limit, we must consider the two cases separately. In the large mass limit we get, at the leading order $1/c^2$, the Schr\"{o}dinger-Newton equation
 \begin{align}
i\hbar\frac{\partial a_0}{\partial t} &= -\frac{\hbar^2}{2m} \nabla^2 a_0 + m \phi({\bf r},t) a_0 \\
\nabla^2\phi(\bf{r},t) &= 4\pi G m\, |a_{0}|^2 \equiv 4\pi G \rho(\bf{r},t)\\
i\hbar\frac{\partial a_0}{\partial t} &= -\frac{\hbar^2}{2m} \nabla^2 a_0  - G m^2\,\int \frac{|a_{0}(\bf{r}^{'}, t)|^2}{|\bf{r} - \bf{r}^{'}|}d^3{\bf r}^{'}\; a_0
\end{align}
where $g_{00}= 1+2\phi/c^2$ defines the Newtonian gravitational potential $\phi$. The Schr\"odinger-Newton equation is known to provide a  spatial localization of wave-packets of large masses because of the gravitational self-interaction: a gravitationally induced inhibition of dispersion \cite{Giulini}. Hence, for $m\gg m_{Pl}$ one is justified in replacing the probability density $|a_0|^2$ by a spatial delta function in the above Poisson equation, giving rise to the classical limit we alluded to below Eqn. (\ref{actualaction3}). Thus the ECD equations have a built in mechanism for macroscopic localization because gravity becomes stronger with increasing mass.

In the small mass limit however, since $L_{CS}\sim \hbar/mc$ goes as $1/c$ [instead of as $1/c^2$ in the large mass limit], we get the surprising result that at orders $1/c$ and $1/c^2$
\begin{equation}
\nabla^2\phi({\bf r},t) =0; \qquad a_0 =0
\end{equation}
For $m\ll m_{Pl}$ the gravitational field as well as the quantum state vanish at order $1/c$. This is an in principle falsifiable prediction of the idea of a unified length $L_{CS}$. An experimental test of the inverse square law for a pair of very small masses would show gravity to be weaker than what is predicted by general relativity.

We notice from the above that curvature vanishes in the small mass limit, whereas in the large mass limit it is torsion that vanishes, because GR holds in the large mass limit. This motivates us to ask if this kind of curvature - torsion duality could be generic. Indeed, we have remarked earlier that since $L_{CS}$ is the only coupling constant in the theory, it will label a large mass solution of the field equations, and also label its dual small mass solution. However, we expect the large mass solution to be gravity dominated, and the small mass solution to be torsion dominated. This is possible only if for a given $L_{CS}$ there are two solutions, one that is curvature dominated, and another that is torsion dominated. We call this the curvature - torsion duality, and construct it as follows. The total curvature $R$ on a space-time manifold can be written as a sum of the Reimannean 
  curvature $\overset{0}{R}$ made from the symmetric Levi-Civita connection, and an additional contribution $Q$ because of torsion:
  \begin{equation}
  R^{\rho}_{\theta\mu\nu} = \overset{0}{R^{\rho}}_{\theta\mu\nu} + Q^{\rho}_{\theta\mu\nu}
  \end{equation}
where
\begin{equation}
Q^{\rho}_{\theta\mu\nu} = \nabla^{\{\}}_\mu K^{\rho}_{\theta\nu} - \nabla^{\{\}}_\nu K^{\rho}_{\theta\mu} + K^{\sigma}_{\theta\nu} K^{\rho}_{\sigma\mu} - K^{\sigma}_{\theta\mu} K^{\rho}_{\sigma\nu}
\end{equation}
$K$ is the contortion tensor and the covariant derivative is with respect to the Levi-Civita connection.

Suppose we have a curvature dominated large mass solution $S1$ with a given $L_{CS}$ and the set of curvature parameters
$[R(1),\overset{0}{R}(1), Q(1)]$. We define the dual torsion dominated solution $S2$ having the same $L_{CS}$ and the set of
curvature parameters $[R(2), \overset{0}{R}(2),Q(2)]$ by the following map:
\begin{equation}
R(1) - Q (1)  = Q (2) - R(2)
\end{equation}
which means that the excess of curvature over torsion for $S1$ equals the excess of torsion over curvature for $S2$. This duality 
implies that $\overset{0}{R}(1)= - \overset{0}{R}(2)$. In the large mass limit, $Q(1)$ is zero and we have the pure curvature solution $R(1) = \overset{0}{R}(1)$. This is general relativity. In the small mass limit, $R(2)$ is zero, and we have the solution
$Q(2) = - \overset{0}{R}(2)$. Since $R(2)$ is zero, this is teleparallel gravity, and the duality map implies that $R(1) = Q(2)$. This duality provides an intriguing connection between GR, ECSK theory, and teleparallel gravity. The first and third theories are respectively the large mass and small mass limit of the ECSK theory and are connected by a duality. We have provided a symmetry between curvature and torsion, and provided a physical basis for Poincar\'e gauge gravity.

This is qualitatively depicted in Fig. 1 where $R-Q$ is plotted against $z\equiv \ln(m/m_{Pl})$. The dual masses $M$ and $m$ have the same value of $L_{CS}$, and the curvature dominated solution $S1$ in the first quadrant is mapped to the torsion dominated solution $S2$ in the third quadrant. As the mass is reduced, a solution `rolls down' from the first quadrant to the origin $m_{Pl}$ and transits to the solution set in the third quadrant. There is also a `mirror universe' whose significance remains to be investigated: For a given $L_{CS}$ the curvature dominated large mass solution is also realized for the dual small mass. This provides the mirror solution which rolls down from the second to the fourth quadrant, and where small masses are curvature dominated, while large masses are torsion dominated. At the transition point $m=m_{Pl}$ we have $R-Q=0$ so that $\overset{0}{R}=0$: this is a Minkowski flat space-time where the total curvature is sourced only by torsion.
\begin{figure} [ht]
 { \centerline{\includegraphics[scale=0.6]{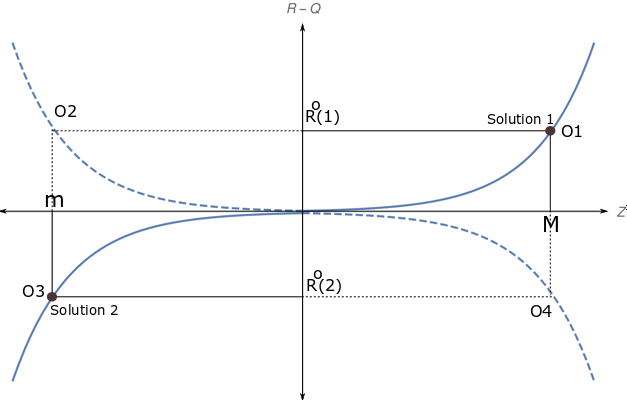}}}
\caption{The curvature - torsion duality }
\end{figure}%

In the small mass limit, where total curvature is zero, torsion balances Reimann curvature. Instead of the Dirac equation, we now have a very special Hehl-Datta equation (\ref{TPS3}), and additonally, as follows from (\ref{TPS1}), $\Sigma^{ij}=0$, so that mass-energy is balanced by spin. We propose this as a starting point for second quantization of the Dirac field, so that one may
investigate if the non-vanishing torsion and Reimann curvature can avoid the infinites of quantum field theory. 

One might ask if it is justified to couple the quantum mechanical Dirac field with classical curvature. The answer is that the Dirac field is quantum only when curvature and torsion are neglected. Soon as coupling to gravity is included, its purely quantum nature is lost, and the quantum to classical transition begins! Furthermore, as we saw, the coupling constant $G$ arises only when the mass becomes large, as if to suggest that gravity emerges only when sufficient mass has accumulated, and then gravity is strictly classical. Only torsion has to be second quantized, which makes sense, because torsion is explicitly expressed in terms of the Dirac state.

The duality map $m\leftrightarrow m_{Pl}^2 / 4m$ can equivalently be thought of as the map $\hbar/2 \leftrightarrow 2Gm^2/c$ from the quantum spin parameter
$\hbar/2$ to the gravitational spin parameter $2Gm^2/c$. Their ratio is the dimensionless gravitational fine structure parameter $4Gm^2/\hbar c$. When this ratio is much larger than one, curvature dominates torsion, and when this ratio is much smaller than one, torsion dominates curvature. The transition from torsion dominance to curvature dominance is marked by the transition from quantum spin to gravitational spin. The suggested relevance of torsion for the Dirac equation opens up a new avenue for research into the relation between space-time geometry and particle physics, at energy scales accessible by current technology.

 \end{document}